\documentclass[prb,twocolumn,amsmath,amssymb,superscriptaddress]{revtex4}

\usepackage{graphicx}
\usepackage{dcolumn}
\usepackage{bm}

\usepackage{float}

\usepackage{amsmath}

\usepackage{epstopdf}

\begin{document}


\title{Anomalous low-energy phonons in nearly tetragonal BiFeO$_3$ thin films}

\author{K.-Y. Choi}
\affiliation{Department of Physics, Chung-Ang University,  Seoul 156-756, Republic of Korea}

\author{S. H. Do}
\affiliation{Department of Physics, Chung-Ang University, Seoul 156-756, Republic of Korea}

\author{P. Lemmens}
\affiliation{Institute for Condensed Matter Physics, TU Braunschweig, D-38106 Braunschweig, Germany}

\author{D. Wulferding}
\affiliation{Institute for Condensed Matter Physics, TU Braunschweig, D-38106 Braunschweig, Germany}

\author{C. S. Woo}
\affiliation{Department of Physics, KAIST, Daejeon 305-701, Republic of Korea}

\author{J. H. Lee}
\affiliation{Department of Physics, KAIST, Daejeon 305-701, Republic of Korea}

\author{K. Chu}
\affiliation{Department of Physics, KAIST, Daejeon 305-701, Republic of Korea}

\author{C.-H. Yang}
\affiliation{Department of Physics, KAIST, Daejeon 305-701, Republic of Korea}

\bigskip
\bigskip
\bigskip

\begin{abstract}
\bigskip
We present evidence for a concomitant structural and ferroelectric transformation around $T_S\sim 360$~K~ in multiferroic BiFeO$_3$/LaAlO$_3$ thin films close to the tetragonal phase. Phonon excitations are investigated by using Raman scattering as a function of temperature. The low-energy phonon modes at 180-260~cm$^{-1}$ related to the FeO$_6$ octahedron tilting  show anomalous behaviors upon cooling through $T_S$; (i) a large hardening amounting to 15~cm$^{-1}$, (ii) an increase of intensity by one order of magnitude, and (iii) an appearance of a dozen new modes. In contrast, the high-frequency modes exhibit only weak anomalies. This suggests an intimate coupling of octahedron tilting to ferroelectricity leading to a simultaneous change of structural and ferroelectric properties.

\end{abstract}


\maketitle


The multiferroic compound BiFeO$_3$ (BFO) is the focus of experimental and theoretical research directed towards room-temperature multifunctional devices~\cite{Wang,Zeches,Bea,Infante}. The distinct multiferroic properties of BFO rely on the high ferroelectric ($T_c\sim1100$~K) and antiferromagnetic ($T_N \sim 640$ K) transition temperature, as well as their mutual couplings~\cite{Roginska,Kiselev}. Bulk BFO has a rhombohedral structure with R3c symmetry. This low symmetry puts a practical constraint on implementing switching devices.

The possibility of circumventing this difficulty is pursued through strain engineering. With increasing strain above 4~\%, an isosymmetric transition takes place from a rhombohedral(R)- to a tetragonal(T)-like structure. The relative fraction of the T and the R phase is controllable as functions of both a film thickness and an electric field~\cite{Zeches,Mazumdar}. It has been shown that the magnetic transition temperature hardly varies but the ferroelectric Curie temperature is strongly reduced with strain~\cite{Infante}. Octahedron tilting is suggested to be responsible for the anomalous strain dependence of ferroelectricity. X-ray diffraction, M\"{o}ssbauer spectroscopy, and piezoresponse force microscopy studies indicate that the majority T-like phase concomitantly undergo structural, magnetic, and ferroelectric transitions  around $T_S\sim 360$~K~\cite{Infante11}.

Raman spectroscopy can serve as a sensitive, local probe of structural and multiferroic properties. Single crystalline and low-strain BFO film exhibited an exceptional coupling of multimagnon, phonon, and electronic excitations~\cite{Ramirez08,Ramirez09}. Raman scattering on highly strained BFO films showed monoclinic (Cc symmetry) distortions from the T-phase~\cite{Iliev}.
However, a detailed temperature study of these effects is unfortunately missing.

In this study, we report on enormous anomalies of low-frequency FeO$_6$ octahedra rotation modes. The concomitant, drastic change of the phonon parameters through $T_S$ suggests the mutual coupling of structural and multiferroic properties.

\begin{figure}
    \centering
    \includegraphics[scale=.35]{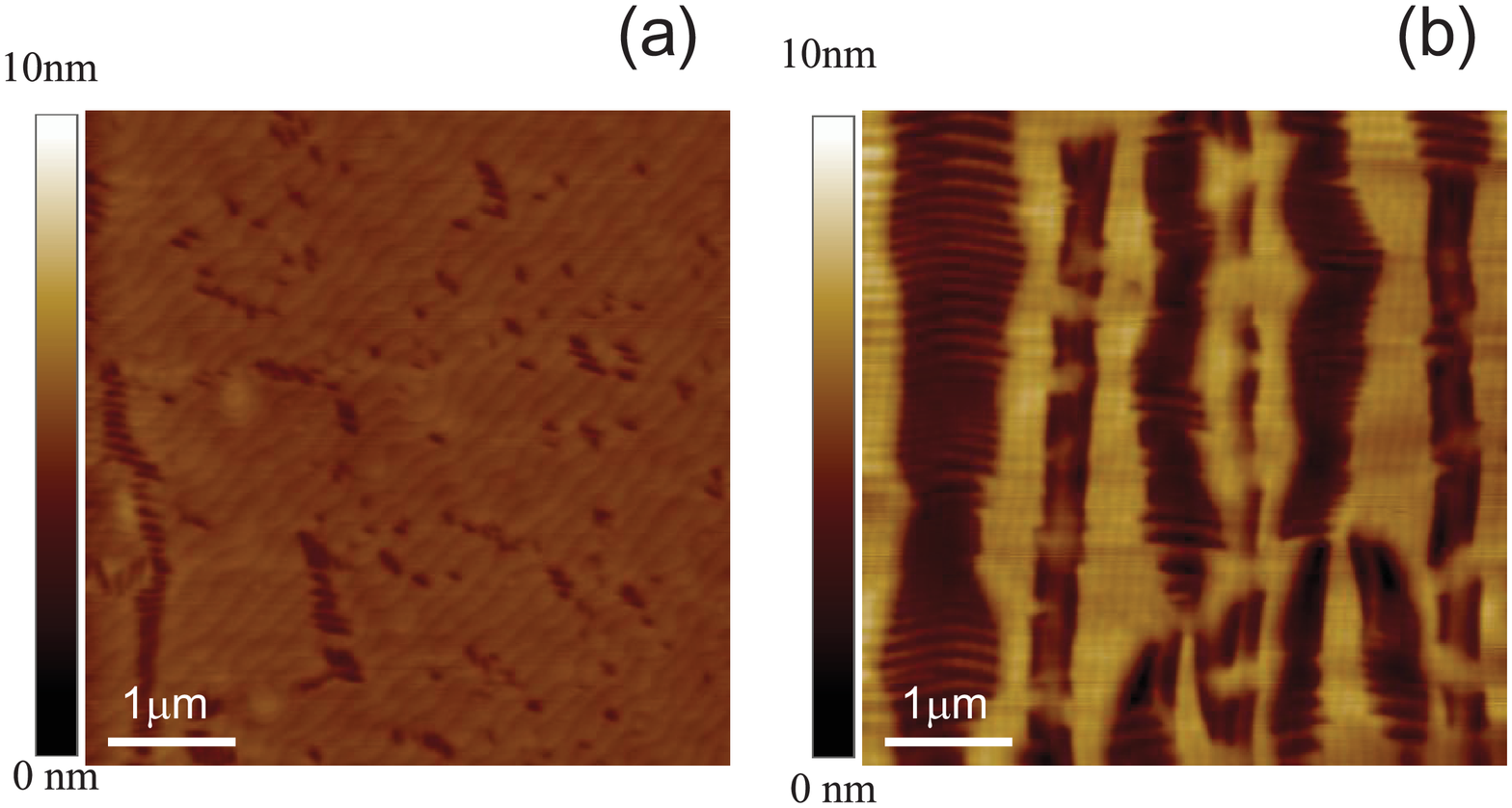}
    \hspace{-22pt}
    \caption{(Color online) Surface morphologies taken by scanning probe microscopy for the BFO films with thickness of (a)~40~nm and (b)~100~nm.}
    \label{AFM}
    \end{figure}

Two BiFeO$_3$ thin films of different thickness (40~nm and 100~nm) were grown on (001) LaAlO$_3$(LAO) substrates by using pulsed laser deposition at the growth temperature of 650$^\circ$C and at the oxygen partial pressure of 100 mTorr. The surface morphologies were investigated by scanning probe microscopy (Veeco Multimode V) with Ti/Pt coated Si tips (MikroMasch). Clear step terrace structures, as shown in Fig.~\ref{AFM}, reveal that both films were grown in a step flow mode, indicating good crystallinity. The thinner film mainly consists of the T-like BFO, while the thicker one contains mixed phase areas (darker areas in the AFM) where the T-like phase and the R phase alternate on nano-length scales. We conclude a strain relaxation in the thicker film. The surface morphologies and the thickness dependence of the phase evolution match well to previous results~\cite{Zeches}.

Raman scattering experiments have been performed in backscattering geometry with the excitation line $\lambda=532$~nm of Nd:YAG solid-state Laser by using a micro-Raman spectrometer (Jobin Yvon LabRam HR).  The light beam was focused to a few $\mu m$-diameter spot on the surface of the BFO thin films using a 50 times magnification microscope objective. The temperature is varied between 10 and 390~K by using a helium cryostat.

    \begin{figure}
	\centering
	\includegraphics[scale=.40]{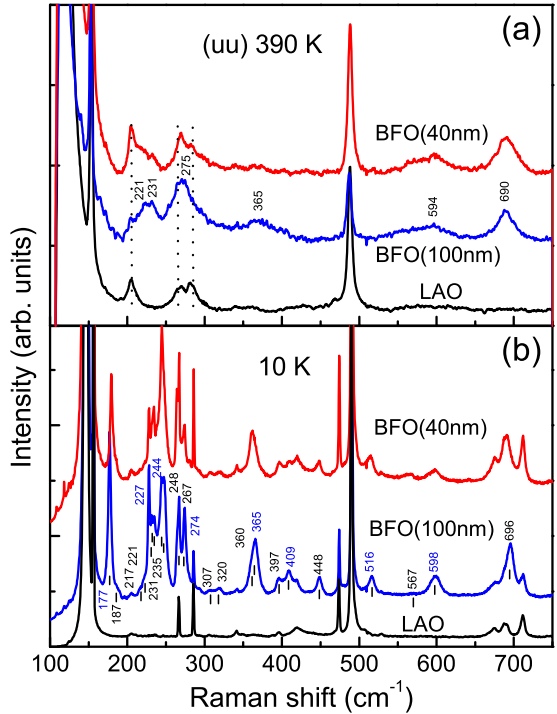}
 	\hspace{-22pt}
	\caption{(Color online) (a) Comparison of unpolarized Raman spectra between BFO~(40nm) and BFO~(100nm) thin films at 390~K. The given numbers are the phonon frequencies of the BFO~(100nm) film. For reference Raman spectrum of a LAO substrate is also shown. (b) Unpolarized Raman spectra
of BFO~(40nm), BFO~(100nm), and LAO at 10~K. The bars denote the peaks from the BFO~(100nm) film.}
	\end{figure}

Figure 2 compares unpolarized Raman spectra of BFO~(40nm) and BFO~(100nm) thin films measured at T=10 K and T=395~K. Due to the small thickness of the studied films, we observe Raman signals from both the substrate and the BFO film. The substrate peaks are identified by measuring separately the LAO crystal. At 390 K we observe 6 phonon modes at 221, 231, 275, 365, 594 and 690~cm$^{-1}$.

For the P4mm tetragonal structure, the factor group analysis yields 8 Raman-active modes of 3A$_{1}$+B$_{1}$+4E. In the backscattering geometry normal to the (001) surface, 3A$_1$+B$_1$ modes of them are symmetry allowed. In comparison to Ref.~\cite{Iliev}, the 221, 275, 365, and 690~cm$^{-1}$ modes are assigned to the tetragonal phase. The two extra modes might be due to either monoclinic distortions or the R-phase.  This indicates that for temperatures above  $T_S\sim 360$~K~\ the BFO film ($t\leq 100$~nm) has a nearly tetragonal structure. At 10~K we identify a total of 22 phonon modes for frequencies above 170~cm$^{-1}$, which match well with 27 expected modes ($\Gamma=14A'+13A''$) for the monoclinic (Cc) symmetry.
This suggests that a more monoclinic-like phase is stabilized at low temperatures~\cite{Xu,Hatt,Iliev}. Both the BFO(40~nm) and the BFO(100~nm) films show the same number of phonons and no noticeable shift in the phonon frequency. The distinct difference is seen in the relative intensities of the 177, 227, 244, 274, 365, 409, 516 and 598~cm$^{-1}$ modes, which are enhanced in the BFO(100~nm) film.  It is noteworthy that the numbers and frequencies  are very close
to those obtained from a rhombohedral crystal structure (compare to Table I of Ref.~\cite{Palai10}).
Since the substrate strain is released with increasing thickness, these modes are attributed to the minority R-phase. This evidences the coexistence of the T-like and the R-like phase.

    \begin{figure}
	\centering
	\includegraphics[scale=.55]{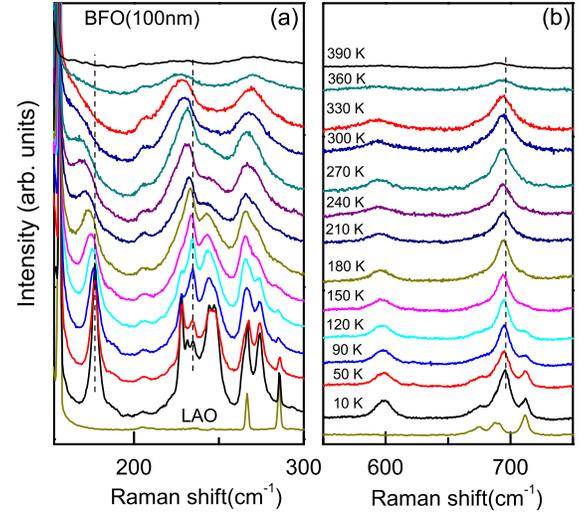}
 	\hspace{-12pt}
	\caption{(Color online)(a) Temperature dependence of the low-frequency Raman spectra of the BFO(100 nm) thin film. The dashed vertical lines are guide to the evolution of phonon frequencies.
 (b)  Temperature dependence of the high-frequency part of Raman spectra.}
	\end{figure}

Figures 3~(a) and (b) zoom into the low- and the high-energy part of the
Raman spectra.  The low-energy modes below 300~cm$^{-1}$  show a drastic temperature dependence in their number, frequency, and intensity while the high-energy modes exhibit a moderate change. To investigate the evolution of the phonon modes in detail we fit them to Lorentzian profiles [see Fig.~5(a)]. The resulting frequency, linewidth, and intensity are plotted as a function of temperature for the two lines at 180 and 690~cm$^{-1}$ in Fig.~4.

  \begin{figure}
	\centering
	\includegraphics[scale=.45]{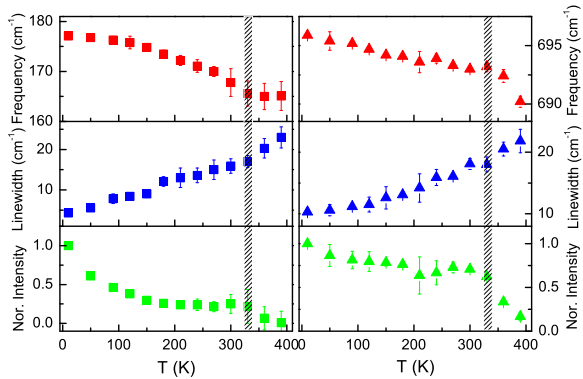}
 	\hspace{-12pt}
	\caption{(Color online)(Left panel)  Temperature dependence of the frequency, the linewidth, and the normalized intensity of the
   low-energy 180~cm$^{-1}$ mode for the BFO(100nm) film. (Right panel)  Temperature dependence of the phonon parameters of the high-energy 690~cm$^{-1}$ mode. The shred bars indicate
   the structural transition temperature.}
	\end{figure}

With lowering temperature the 180-cm$^{-1}$ mode undergoes a large hardening  by 15~cm$^{-1}$, starting at
around $T_S$. Its linewidth decreases monotonically and its scattering intensity grows by one order of magnitude. As the Raman scattering intensity is proportional to the square of the derivative of the dielectric function with respect to the amplitude of the normal mode, the strong intensity increase means that the low-energy modes are susceptible to a change of ferroelectricity. Since the low-energy modes are associated with the external vibrations of the FeO$_6$ octahedra~\cite{Choi}  and polar cation displacements~\cite{Rovillain}, they provide a measure of the tilting degree of the FeO$_6$ octahedra~\cite{KY}. In this light, the concomitant change of intensity and frequency suggests that the increase of octahedral tilting accompanies the polar cation displacements. Exactly, this cross-coupling effect has been discussed as an origin of the reduction of the ferroelectric transition temperature under strain~\cite{Infante}.

The high-frequency modes are related to the internal vibrations of the FeO$_6$ octahedra and are susceptible to a change in the Fe-O bond distance and angle. The phonon parameters of the 690~cm$^{-1}$ mode show only small anomalies at about T$_S$.  Since the Fe-O bond angle and length determines a strength of superexchange interactions, we conclude that the magnetic interactions are rather weakly coupled to ferroelectricity.

 \begin{figure}[t]
	\centering
	\includegraphics[scale=.50]{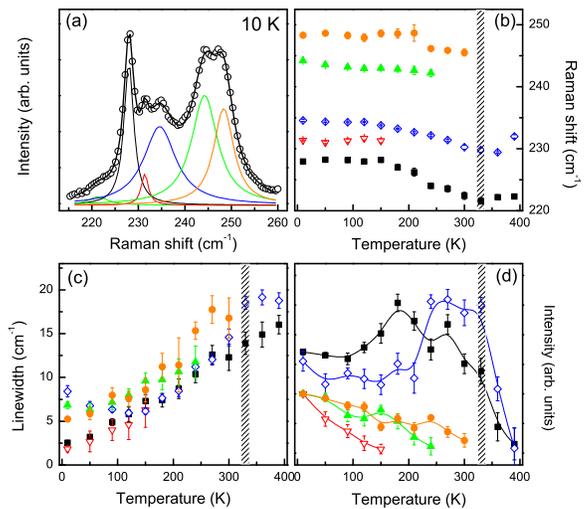}
 	\hspace{-12pt}
	\caption{(Color online) (a) Fit of the low-energy Raman spectrum  to Lorentzian profiles at 10~K.
 (b), (c), and (d)  Temperature dependence of the optical phonon frequencies, linewidths and intensities.}
	\end{figure}

In Fig.~5 we focus on the evolution of the low-frequency phonons with $\Delta\omega=220 - 260~\mbox{cm}^{-1}$. With decreasing temperature, several anomalies develop.
For temperatures below $T_S$, several extra modes appear at 231, 244, and 248
cm$^{-1}$.  Between $T_S$ and 150~K  the phonon frequencies
undergo a substantial hardening and the linewidths strongly narrow.
Noteworthy is the large enhancement of the 222- and 234-cm$^{-1}$ modes upon
cooling through $T_S$ and the small drop of their intensity for temperatures
below 170 - 240~K. The concomitant, drastic change of the phonon number and intensity
indicates a complex nature of the phase transition. The increased number of the
phonon modes is due to a transition from the more T-like phase to the monoclinic-like
one.~\cite{Infante11}  The structural change can also lead to the enhancements of phonon scattering intensity. However, the observed huge anomalies invoke the strong change of electronic polarizabilities beyond the structural transformation. As discussed in the $180~\mbox{cm}^{-1}$ rotation mode, the octahedral tiltings are intimately coupled
to ferroelectricity. The increase of the octahedral tilting degree seems to be
closely related to the stabilization of the monoclinic-like phase. Therefore,
the intensity-polarizability enhancements of the low-energy
modes are evidence for the coupled structural-ferroelectric transition.
Actually, Infante {\it et al.,}~\cite{Infante11} showed that a hard to a soft
ferroelectric transition accompanies the structural and magnetic phase transitions.

In summary, we have presented a Raman scattering study of nearly tetragonal BFO thin films in the temperature range of 10 - 390~K. We observe large anomalies of the internal phonon modes of $180-300 \mbox{cm}^{-1}$, which are susceptible to the tiltings of the FeO$_6$ octahedron. The most salient feature is the strong enhancement of phonon intensity and the increase of a phonon number through a structural phase transition. Our study suggests that a structural and ferroelectric transition is closely tied by cross-coupling effects between octahedron tiltings and polar cation displacements.

We acknowledge financial support from DFG, the Alexander-von-Humboldt Foundation and Korea NRF Grants No. 2009-0093817 and No. 2010-0013528.


\begin{thebibliography}{10}

\bibitem{Wang}
J. Wang, J. B. Neaton, H. Zheng, V. Nagarajan, S. B. Ogale, B. Liu, D. Viehland, V. Vaithyanathan, D. G. Schlom, U. V. Waghmare, N. A. Spaldin, K. M. Rabe, M. Wutting, R. Ramesh, Science \textbf{299}, 1719 (2003).

\bibitem{Zeches}
R. J. Zeches, M. D. Rossell, J. X. Zhang, A. J. Hatt, Q. He, C.-H. Yang, A. Kumar,
C. H. Wang, A. Melville, C. Adamo, G. Sheng, Y.-H. Chu, J. F. Ihlefeld, R. Erni, C. Ederer,
V. Gopalan, L. Q. Chen, D. G. Schlom, N. A. Spaldin, L. W. Martin, and R. Ramesh, Science \textbf{326}, 977 (2009).

\bibitem{Bea}
H. Bea, B. Dupe, S. Fusil, R. Mattana, E. Jacquet, B. Warot-Fonrose, F. Wilhelm, A. Rogalev,
S. Petit, V. Cros, A. Anane, F. Petroff, K. Bouzehouane, G. Geneste, B. Dkhil, S. Lisenkov, I. Ponomareva, L. Bellaihe, M. Bibes, A. Barthelemy, Phys. Rev. Lett.  \textbf{102}, 217603 (2009).

\bibitem{Infante}
I. C. Infante, S. Lisenkov, B. Dup\'{e}, M. Bibes, S. Fusil, E. Jacquet, G. Geneste, S. Petit,
A. Courtial, J. Juraszek, L. Bellaiche, A. Barth\'{e}l\'{e}my, and B. Dkhil, Phys. Rev. Lett.  \textbf{105}, 057601 (2010).


\bibitem{Roginska}
Y. E. Roginska, Y. Y. Tomashpo, Y. N. Venevtse, V. M. Petrov, and G. S. Zhdanov, Sov. Phys. JETP \textbf{23}, 47-51 (1966).

\bibitem{Kiselev}
S. V. Kiselev, R. P. Ozerov, and G. S. Zhdanov, Sov. Phys. Dokl. \textbf{7}, 742-744 (1963).



\bibitem{Mazumdar}
D. Mazumdar, V. Shelke, M. Iliev, S. Jesse, A. Kumar, S. V. Kalinin, A. P. Baddorf,
and A. Gupta, Nano Lett.  \textbf{10}, 2555 (2010).


\bibitem{Infante11}
I. C. Infante, J. Juraszek, S. Fusil, B. Dup\'{e}, P. Gemeiner, O. Di\'{e}guez, F. Paillouz, S. Jouen,
 E. Jacquet, G. Geneste, J. Pacaud, J. \'{I}niguez, L. Bellaiche, A. Barth\'{e}l\'{e}my, B. Dkhil,
 and M. Bibes, arXiv: 1105.6016 (2011), unpublshed.



\bibitem{Ramirez08} M. O. Ramirez, M. Krishnamurthi, S. Denev, A. Kumar, S-Y. Yang, Y-H. Chu, E. Saiz, J. Seidel, A. P. Pyatakov, A. Bush, D. Viehland, J. Orenstein, R. Ramesh, and V. Gopalan, App. Phys. Lett. \textbf{92}, 022511 (2008).

\bibitem{Ramirez09} M. O. Ramirez, A. Kumar, S. A. Denev, Y. H. Chu, J. Seidel, L. W. Martin S.-Y. Yang, R. C. Rai, X. S. Xue, J. F. Ihlefeld, N. J. Podraza, E. Saiz, S. Lee, J. Klug, S. W. Cheong, M. J. Bedzyk, O. Auciello, D. G. Schlom, J. Orenstein, R. Ramesh, J. L. Musfeldt, A. P. Litvinchuk, and V. Gopalan, App. Phys. Lett. \textbf{94}, 161905 (2009).


\bibitem{Iliev}
M. N. Iliev, M. V. Abrashev, D. Mazumdar, V. Shelke, and A. Gupta, Phys. Rev. B \textbf{82}, 014107 (2010).



\bibitem{Xu}
G. Xu, H. Hiraka, G. Shirane, J. Li, J. Wang, and D. Viehland, Appl. Phys. Lett. \textbf{86}, 182905 (2005).

\bibitem{Hatt} A. J. Hatt, N. A. Spaldin, and C. Ederer, Phys. Rev. B \textbf{81}, 054109 (2010).


\bibitem{Palai10}
R. Palai, H. Schmid, J. F. Scott, and R. S. Katiyar, Phys. Rev. B  \textbf{81}, 064110 (2010).


\bibitem{Choi}
Although Fe$^{3+}$ ions of the T-like phase is thought to be coordinated in FeO$_5$ pyramids,
isomer shift values of M\"{o}ssbauer spectroscopy indicate that Fe$^{3+}$ ions of
both the T- and R-like phases are in octahedral enviroments~\cite{Infante11}.



\bibitem{Rovillain}
P. Rovillain, M. Cazayous, Y. Gallais, A. Sacuto, R. P. S. M. Lobo, D. Lebeugle,
and D. Colson, Phys. Rev. B  \textbf{79}, 180411(R) (2009).




\bibitem{KY} K.-Y. Choi, P. Lemmens, T. Sahaoui, G. Guntherodt, Yu. G. Pashkevich, V. P. Gnezdilov, P. Reutler, L. Pinsard-Gaudart, B. Buchner, and A. Revcolevschi,
Phys. Rev. B \textbf{71}, 174402 (2005).





\end{thebibliography}
\end{document}